\documentclass[aps,prapplied,reprint,superscriptaddress]{revtex4-2}

\usepackage{amsmath,amsfonts,amssymb}
\usepackage{graphicx}
\usepackage{hyperref}

\begin{document}

\title{Quadrature-averaged homodyne detection for cavity parameter estimation}

\author{Giada R. La Gala}
\affiliation{Center for Nanophotonics, AMOLF, Science Park 104, 1098 XG, Amsterdam, The Netherlands}

\author{Arvind Shankar Kumar}
\affiliation{Department of Physics and Nanoscience Center, University of Jyv{\"a}skyl{\"a}, P.O. Box 35, FI-40014 University of Jyv{\"a}skyl{\"a}, Finland}

\author{Rick Leijssen}
\affiliation{Center for Nanophotonics, AMOLF, Science Park 104, 1098 XG, Amsterdam, The Netherlands}

\author{Ewold Verhagen}
\affiliation{Center for Nanophotonics, AMOLF, Science Park 104, 1098 XG, Amsterdam, The Netherlands}

\author{Juha T. Muhonen}
\affiliation{Department of Physics and Nanoscience Center, University of Jyv{\"a}skyl{\"a}, P.O. Box 35, FI-40014 University of Jyv{\"a}skyl{\"a}, Finland}

\begin{abstract}
Balanced homodyne interferometry is a well-known detection technique that allows for sensitive characterization of light fields. Conventionally a homodyne interferometer is operated by locking the relative phase of a reference beam to the signal beam by means of an active feedback loop.
A less often used method is to perform a slow continuous modulation of the reference beam arm length that corresponds to averaging all relative phases during the measurement. 
Here we show theoretically and experimentally that this quadrature averaging can be advantageous in estimating the parameters of a resonant optical cavity.
We demonstrate that the averaging turns the transduction function, from cavity frequency fluctuations into the interferometer signal, into a simple function of the laser detuning that, notably, does not depend on the parameters of possible non-resonant channels present in the system. 
The method needs no active feedback and gives results that are easy to interpret. Moreover, the phase-averaged measurement allows to characterize the absolute magnitude of a cavity frequency modulation.
\end{abstract}

\date{\today}

\maketitle

\section{Introduction} 
Balanced homodyne interferometry (BHI) offers a unique tool to characterize arbitrary quadratures of a light field with photon shot noise limited sensitivity, and is in use in fields varying from detection of gravitational waves \cite{LIGO} to quantum applications of cavity optomechanics \cite{Aspelmeyer2014}.
A common use case for BHI is to probe an optical cavity that is located in one of the arms of the interferometer (called the signal arm). Any frequency changes of that optical cavity are transferred into relative phase differences between the two interferometer arms (signal and local oscillator). Importantly, the induced phase difference is also enhanced by the cavity finesse. 

Usually the detection of the phase difference is performed by locking the local oscillator arm phase with regards to the (non-perturbed) signal phase by the means of a variable path length (piezo mirror, fiber strecher etc.) \cite{bachor_guide_2019}. In this way a particular quadrature of the light field can be measured \cite{lvovsky_continuous-variable_2009}. This obviously assumes that the signal to be measured is at a much higher frequency than the locking bandwidth in order for the feedback not to cancel out the signal. In addition, the actual angle of the interferometer is generally not an accessible observable and hence ``locking the angle'' is usually accomplished by locking the DC signal of the interferometer output. As we show below, this can actually lead to unwanted side-effects for the measurement if the signal also includes a non-resonant background.

One method used to avoid the effects of the non-resonant channels is to convert the homodyne signal into a ''pseudo-heterodyne'' signal by modulating the local oscillator path with a frequency $\omega_1$ \cite{Jackson1982}. This will create sidebands for the detected signal at frequencies separated from the original detection frequency by $\omega_1$ allowing separating the interferometer signal from the background signals. This method has for example been applied in scanning near-field optical microscopes \cite{Sasaki_2000,Hillenbrand2000,Nenad2006}. In this method $\omega_1$ needs to be much larger than the measurement bandwidth around the signal to be detected so that well-defined sidebands are created.

A less often used method, that we focus on here, is to average over all possible phases of the local oscillator by modulating the local oscillator slowly, i.e., with a much lower frequency than the bandwidth of the measurement. This modulation will then not create sidebands but rather average over all the possible angles during the homodyne measurement \cite{Munroe1995,mcalister97,Lvovsky2001}. Here we show that this quadrature-averaged BHI can be a useful method in characterizing optical cavities when the measured signal also includes light that has not interacted with the cavity (non-resonant reflection or transmission). This has applications especially in cavity optomechanics \cite{aspelmeyer_cavity_2014}, and in other areas using nanophotonic cavities, where the non-resonant channels can be sensitive to the experimental conditions \cite{Galli2009,miroshnichenko_fano_2010,li_experimental_2011,Ding14,Zhao16,leijssen_strong_2015,leijssen_nonlinear_2017,limonov_fano_2017,naesby_microcavities_2018}. In a conventional locked homodyne measurement the interference with the resonant and non-resonant part generally leads to a Fano shaped resonance from which the parameters of the cavity can be hard to extract. Additionally, the homodyne interferometer angle will vary as a function of the detuning of the laser and the optical cavity. We show that by using quadrature-averaged BHI we can avert both of these problems. A simple Lorentzian shape for the resonance is recovered, allowing easier interpretation, and the variation in the measurement angle with laser detuning is eliminated. Moreover, the quadrature-averaged BHI allows extracting the absolute magnitude of the cavity frequency modulation by comparing the different harmonic components. This can then be used to retrieve e.g. the thermal modulation amplitude of an optomechanical resonator without any further calibration.

\section{Background: Homodyne interferometry in the presence of non-resonant channels}

\subsection*{Homodyne interferometer description}

\begin{figure}
    \centering
    \includegraphics[width=0.3\textwidth]{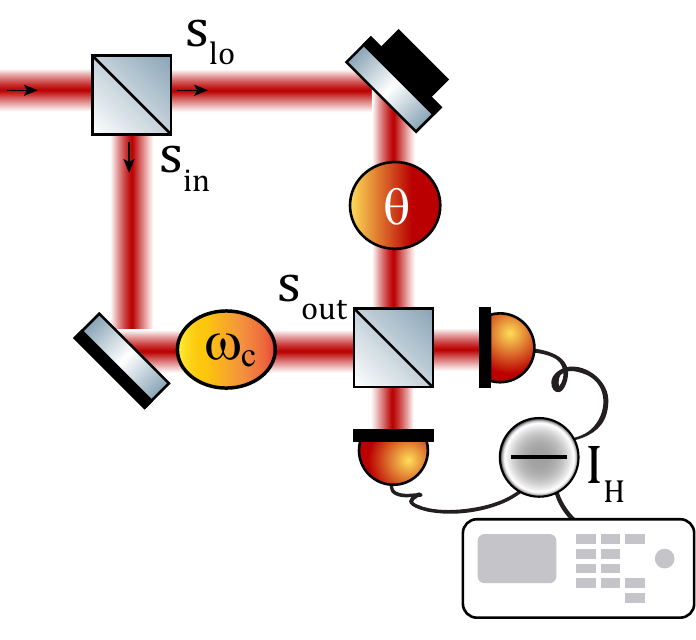}
    \caption{Schematic of a homodyne interferometer. Symbols are defined in the main text.}
    \label{fig:homodyne}
\end{figure}

In figure~\ref{fig:homodyne} we show a schematic of a balanced homodyne interferometer. We assume that the signal and the local oscillator arms are described by linearly polarized coherent states with complex parameters $s_\mathrm{in}= \left|s_\mathrm{in}\right|e^{i\phi_\mathrm{in}}$ and $s_\mathrm{lo}=\left|s_\mathrm{lo}\right|e^{i\phi_\mathrm{lo}}$. By changing the local oscillator arm length, the phase $\phi_\mathrm{lo}$ can be tuned and a relative phase difference $\theta$ is imprinted between the two paths. By convention, we normalize all fields $s_i$ such that $|s_i|^2$ is the photon flux in beam $i$ and hence the power in the beam is $P_i = \hbar\omega_l|s_i|^2$, where $\omega_l$ is the frequency of the light. The optical cavity under study (with resonance frequency $\omega_c$) is placed on the signal arm and due to interaction with the cavity system, the signal beam $s_\mathrm{in}$ is transformed into the output beam $s_\mathrm{out}=r s_\mathrm{in}$, where $r$ is the \textit{transfer function} describing the cavity system. Subsequently the two beams are interfered at the final 50:50 beam splitter leading to outputs $s_+=1/\sqrt{2}(i s_\mathrm{out} + s_\mathrm{lo})$ and $s_-=1/\sqrt{2}( i s_\mathrm{lo} + s_\mathrm{out} )$ which are detected by separate photodiodes. The two output currents are then subtracted. The measured subtracted photocurrent from the detector ($I_H$) can be written as
\begin{eqnarray}
I_H &=& \left|s_\mathrm{+}\right|^2-\left|s_\mathrm{-}\right|^2 = s_+^\star s_+ - s_-^\star s_- \nonumber \\
&=& i (s_\mathrm{lo}^\star s_\mathrm{out} - s_\mathrm{out}^\star s_\mathrm{lo}) \nonumber \\
&=&-2\left|s_\mathrm{in}\right|\left|s_\mathrm{lo}\right| \left( \Im{(r)} \cos\theta - \Re{(r)}\sin\theta \right),
\label{eq:homodyneresponse}
\end{eqnarray}
where $\theta=\phi_\mathrm{lo}-\phi_\mathrm{in}$ is the phase difference between the two arms of the interferometer and $\Re{(r)}$ and $\Im{(r)}$ are the real and imaginary parts of the cavity transfer function, respectively. (Note that changing the convention to $\theta=\phi_\mathrm{in}-\phi_\mathrm{lo}$ would change the sign in front of the sine term.)
We have neglected any gain/loss factor depending on the quantum efficiencies of the detectors, the amplifier gain and other electronic properties; these would simply lead to a constant multiplier for this term.
Varying $\theta$ between $0$ and $2\pi$ (by tuning $\phi_\mathrm{lo}$) corresponds to choosing a specific quadrature of the field: where $\theta=0$ and $\theta=\pm\pi$ are the extreme cases usually known as the \textit{phase quadrature} and the \textit{amplitude quadrature}. These are orthogonal quantities in the optical phase space.

\subsection*{Input-Output description of a cavity system}

The beam $s_\mathrm{out}$ carries information about the cavity system. For a single-mode, high-finesse cavity, the transfer function can be described by a general resonant response \cite{haus1984waves}
\begin{eqnarray}
    r&=&ce^{i\varphi} - \frac{\eta\kappa}{\kappa/2+i (\omega_c - \omega)} \nonumber \\
     &=& ce^{i\varphi} - \frac{2\eta}{1+i\Delta_n},
    \label{eq:tf}
\end{eqnarray}
where $\Delta\equiv\omega_c-\omega$ is the detuning of the incident laser frequency $\omega$ from the cavity resonance frequency $\omega_c$ and we have introduced a normalized detuning parameter $\Delta_n = 2\Delta/\kappa$. 
The first term in this general expression describes a non-resonant channel, with amplitude $c$ and phase $\varphi$, e.g. a direct reflection of light that does not enter the cavity. The second term describes the cavity resonance, which is characterized by the resonance frequency $\omega_c$, cavity decay rate $\kappa$ and radiative coupling efficiency $\eta\equiv\kappa_\mathrm{ex}/\kappa$ where $\kappa_\mathrm{ex}$ is both the decay rate through the radiative coupling channel, i.e., radiated into the beam $s_\mathrm{out}$, and the coupling from the input channel to the cavity (assumed equal). We note that in the most general case, the transfer function could still be multiplied by an overall phase factor, but any global phase factor can always be absorbed into a proper adjustment of the phase $\theta$ when evaluating the resulting homodyne signal.

It is also useful to write down the normalized transfer function that is unity at resonance (without the non-resonant part) 
\begin{equation}
r_n \equiv \frac{r}{2\eta} = c_n  e^{i\varphi} - \frac{1}{1+i\Delta_n}, 
\end{equation}
where $c_n=c/(2\eta)$ is the normalised $c$ that will appear in all formulae below. We note that the case $c=1,\varphi=0$, corresponds to the classical time average of the famous input-output expression of $s_\mathrm{out} = s_\mathrm{in}-\sqrt{\kappa_{ex}} s$, where $s$ is the field inside the cavity.

From equation~(\ref{eq:homodyneresponse}) we see that specific homodyne angles ideally probe the real or imaginary part of the cavity response, given by
\begin{eqnarray}
    \Re{(r_n)} &=c_n\cos\varphi - \frac{1}{1+\Delta_n^2}, \label{eq:real} \\
    \Im{(r_n)} &=c_n\sin\varphi + \frac{\Delta_n}{1+\Delta_n^2}. \label{eq:imag}
\end{eqnarray}

In general, depending on the system characteristics (the number of coupling channels, the relative loss rates, relative phases) the resonant lineshapes as a function of $\Delta$ can vary strongly. This can be understood as a Fano interference effect \cite{Fano1961}, as it arises due to the interference of the resonant and the direct (broadband) channel between input and output. This is important for experiments as the amplitude and phase of the non-resonant part can be very sensitive to the experimental conditions in such way that the detuning dependence of a specific quadrature is strongly affected by small changes in the experimental conditions \cite{miroshnichenko_fano_2010,Ding14,Zhao16,leijssen_strong_2015,leijssen_nonlinear_2017}.

\begin{figure*}
    \centering
    \includegraphics[width=0.65\textwidth]{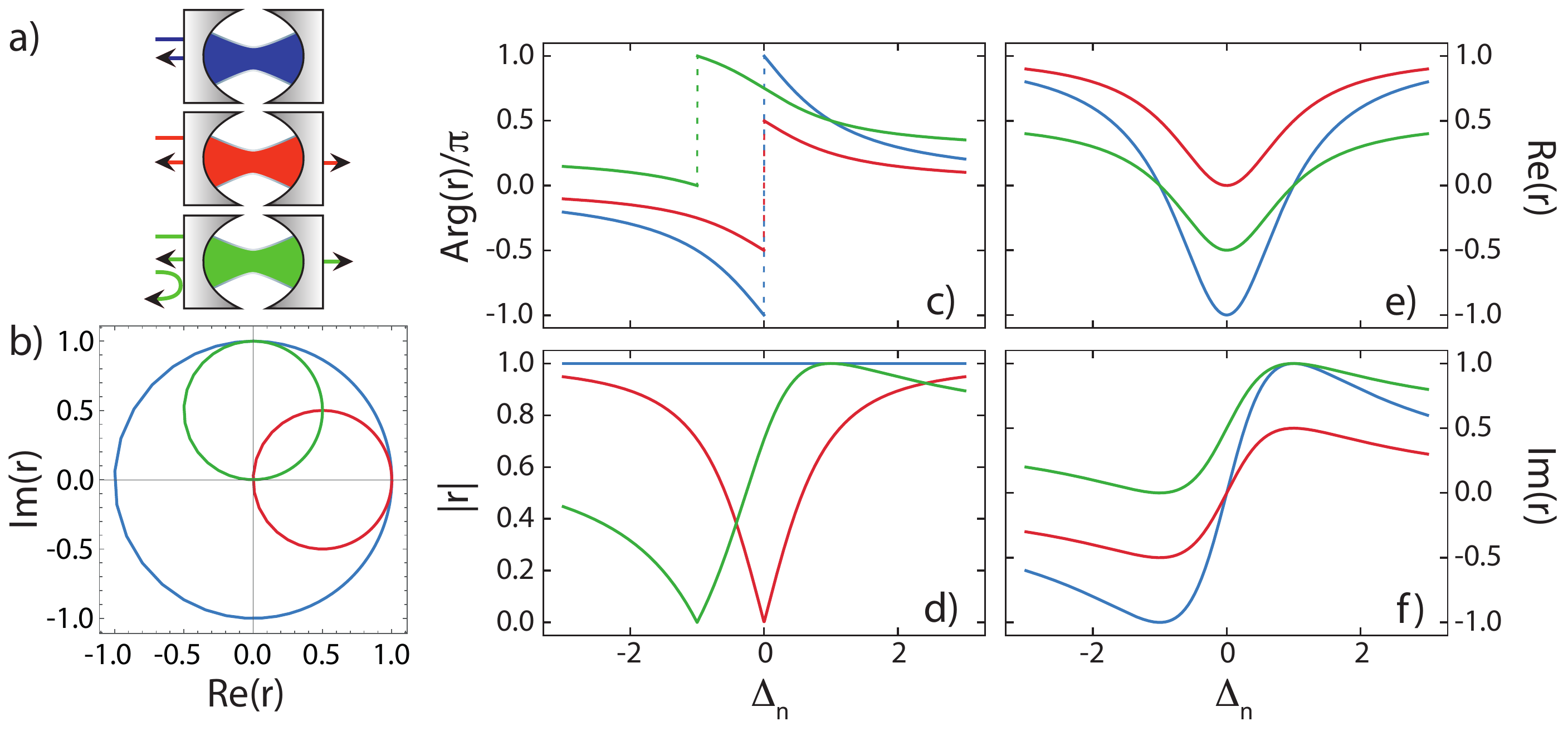}
    \caption{a) Three optical systems, i.e. Fabry-P\'{e}rot cavities with different coupling channels. In blue the case of a single sided cavity with only one resonant channel, in red the case of a double sided cavity with two resonant coupling channels and in green the case of a double sided cavity plus a non-resonant channel. b) The parametric plot shows the transfer function in the complex plane for the three cavities with the same color convention, whereas c-d) are the amplitude and phase of the complex transfer function versus laser detuning and e-f) its real and imaginary parts.}
    \label{fig:fpcavity}
\end{figure*}

To illustrate the effect of the non-resonant channel, we plot in figure~\ref{fig:fpcavity} the phase and amplitude (and the real and imaginary parts separately) of the transfer function as a function of detuning for three particular cases: i) a \textit{single sided} cavity ($c=1$, $\varphi=0$, $\eta=1$), where all the light in the cavity is returned towards the detector, ii) a \textit{double sided} cavity ($c=1$, $\varphi=0$, $\eta=0.5$), where half the light is lost due to some other channels (i.e., critical coupling), and iii) a double sided cavity with also a non-resonant \textit{back reflection} from the first mirror ($c=1/\sqrt{2}$, $\varphi=\pi/4$, $\eta=1/2$). From the phasor diagram (figure~\ref{fig:fpcavity}b) all the plots can be retrieved conveniently, by noticing that, from equation~(\ref{eq:tf}), the zero for $\Delta$ corresponds to the minimum of $\Re{(r)}$ in the phase convention we took in equation~(\ref{eq:tf}).

The resonance parameters  $c, \varphi, \eta$ are bounded and related to each other by conservation of energy, which restricts the sum of the intensities at the output of all decay channels to be the same as the intensity at the input. For our formalism, this means that $|r|^2 \leq 1$. We allow $|r|$ to be less than one as we are considering also the presence of other channels that do not radiate to $s_\mathrm{out}$. In general one obtains a Fano lineshape for the reflected amplitude $|r|$, which can vary from a dispersive to absorptive (Lorentzian) shape depending on the value of $\varphi$. The Lorentzian lineshape is characteristic of relative phases between resonant and non-resonant channels that are 0 or $\pi$. The standard double sided Fabry-Perot cavity is an example, which has full transmission on resonance and large reflectance for large detuning. But in more complex systems also dispersive lineshapes can be encountered.

\subsection*{Measuring fluctuations: time dependent output}
We now demonstrate that, in the presence of a non-resonant channel, described by the first term on the rhs of equation~(\ref{eq:tf}), the measurement of cavity frequency fluctuations by means of a standard homodyne detection scheme is affected by the background contribution, represented in the equations by the dependence of the output signal from the parameters $c$ and $\varphi$. This fact generally complicates the analysis of experimental data. We will then present a method for eliminating this problem.

The effect of small modulation of cavity resonance frequency $\delta\omega_c$ is translated into the measured signal in the form of fluctuations of the homodyne output.
In the presence of a modulation of the cavity frequency, the laser detuning $\Delta$ becomes time-dependent
\begin{equation}
    \Delta(t)=\bar{\Delta}+\delta\Delta(t)=\omega-\bar{\omega}_c-\delta\omega_c(t),
\end{equation}
where $\delta\Delta$ and $\delta\omega_c$ denote the detuning and cavity resonance frequency fluctuations, and the horizontal bars denote mean values.
In order to address mechanical fluctuations we want to measure the fluctuations of the homodyne signal $I_H(t)=\bar{I}_H+\delta I_H(t)$. 
The general expression for the output signal $I_\mathrm{H}(t)$ from equations (\ref{eq:homodyneresponse}) and (\ref{eq:tf}) is 
\begin{eqnarray}
       I_\mathrm{H}&=& 4 \eta \left|s_\mathrm{in} \right| \left|s_\mathrm{lo} \right| 
      \biggl[ c_n \sin(\varphi-\theta) \nonumber \\ 
      && +\left. \frac{1}{1+\Delta_n^2}\left(\Delta_n \cos\theta + \sin\theta\right) \right].
\label{eq:deltaph}
\end{eqnarray}

We are for now interested in small fluctuations and hence we linearize the response with respect to changes in the detuning parameter $\Delta$, $\delta I_H = \frac{dI_H}{d\Delta} |_{\Delta_n = \bar{\Delta}_n} \times \delta \omega_c$ it follows that
\begin{eqnarray}
        \delta I_\mathrm{H}(t)&=& \frac{\delta\omega_c}{\kappa} \left|s_\mathrm{in} \right| \left|s_\mathrm{lo} \right| \frac{8\eta}{\left(1+\bar{\Delta}_n^2\right)^2} \nonumber \\
        && \times \left[\left(1-\bar{\Delta}_n^2\right)\cos\theta -2\bar{\Delta}_n\sin\theta\right] \label{eq:ph_fluc} \\
        &\equiv& \frac{\delta\omega_c}{\kappa} \left|s_\mathrm{in} \right| \left|s_\mathrm{lo} \right| \beta(\bar{\Delta}_n,\eta,\theta), \nonumber
\end{eqnarray}
where we have defined the \textit{transduction function} 
\begin{equation}
\beta = \frac{8\eta}{\left(1+\bar{\Delta}_n^2\right)^2}
        \left[\left(1-\bar{\Delta}_n^2\right)\cos\theta -2\bar{\Delta}_n\sin\theta\right]
\end{equation}
that defines how (normalized) frequency fluctuations in the cavity frequency are transduced into the detected homodyne signal (assuming linearization). 

In order to access a particular quadrature of the signal $s_\mathrm{out}$, ideally one would fix a value for the phase delay $\theta$. In this case, the fluctuating signal $\delta I_H (t)$ would be independent on the background parameters $c_n,\varphi$, as one can see from equation~(\ref{eq:ph_fluc}). However, experimentally $\theta$ is not an accessible parameter and cannot be fixed directly. The measurement is accomplished, instead, by fixing the average output signal $\bar{I}_H$ to a constant value by introducing a feedback loop on the reference beam length (and hence phase $\phi_{lo}$). The dramatic drawback consist in having $\delta I_H$ dependent on the parameter $c$ and $\varphi$ through the phase $\theta$.

For example, if we keep $\bar{I}_H=0$ as is most conventionally done, we have
\begin{equation}
\label{eq:phase}
    \theta(\bar{\Delta}_n)=\arctan\left(-\frac{\Im{(r)}(\bar{\Delta}_n)}{\Re{(r)}(\bar{\Delta}_n)}\right) + 2\pi n,
\end{equation}
where $n$ is an integer. By using (\ref{eq:real}) and (\ref{eq:imag}) the ratio has an analytic form
\[-\frac{\Im{(r)}}{\Re{(r)}} = \frac{\bar{\Delta}_n+(1+\bar{\Delta}_n^2)c_n\sin\varphi}{1-(1+\bar{\Delta}_n^2)c_n\cos\varphi},  \]
and hence we have that (modulo 2$\pi$)
\[\theta=\theta(\bar{\Delta}_n,c_n,\varphi) = \arctan\left(  \frac{\bar{\Delta}_n+(1+\bar{\Delta}_n^2)c_n\sin\varphi}{1-(1+\bar{\Delta}_n^2)c_n\cos\varphi} \right). \]

If instead one would want to maximize the difference between the balanced detectors such that $\partial\bar{P}_H/\partial\theta=\Re{(r)}(\bar{\Delta})\cos\theta-\Im{(r)}(\bar{\Delta}_n)\sin\theta=0$ one would obtain
 \begin{equation}
 \label{eq:amplitude}
    \theta(\bar{\Delta}_n)=\arctan\left( \frac{\Re{(r)}(\bar{\Delta}_n)}{\Im{(r)}(\bar{\Delta}_n)} \right) +2\pi n.
\end{equation}
This can be again written as above using equations (\ref{eq:real}) and (\ref{eq:imag}).

Hence using this scheme makes the homodyne angle $\theta$ dependent on the non-resonant parameters  $\theta=\theta(\bar{\Delta}_n,c_n,\varphi)$. These expressions also vary with $\bar{\Delta}_n$ and as a result these settings do not necessarily probe the phase or amplitude quadrature for all laser frequencies. Moreover, if $c$ and $\varphi$ are not precisely known, even at detuning $\bar{\Delta}=0$ the probed quadrature is not determined.
As an example, consider the usual case of $\bar{I}_H=0$. Then if $c=0$, we have
$\theta = \arctan \left(\Delta_n\right)$, which will probe the wanted quadrature $\theta=0$, when $\Delta = 0$. (Hence, even in this ideal case we still need to know the detuning parameter separately.) However, taking into account the non-resonant channel means that \text{even if we can separately set $\Delta=0$} we will end up with $\theta = \arctan \left(c_n\sin\varphi /(1-c_n\cos\varphi)\right)$, which means that the transduction parameter will depend on the non-resonant channel. 

All the above is not really a problem for the clear-cut cases where $c,\varphi$ are known. However, especially for light interaction with complex nanophotonic systems such as photonic crystal cavities, to make quantitative predictions about the non-resonant reflection one should acquire knowledge about the optical setup configuration such as optical beam size and relative polarization between optical cavity mode and signal beam polarization~\cite{Galli2009}. The values of $c$ and $\varphi$ can vary from sample to sample and depend strongly on optical alignment. In fact, in the optomechanical devices based on photonic crystal cavities it is often impossible to measure $c$ and $\varphi$ as the reflectance $|r|$ as shown in figure~\ref{fig:fpcavity} cannot be directly observed because of large fluctuations and/or (optomechanical) nonlinearities \cite{leijssen_strong_2015,leijssen_nonlinear_2017}.

\subsection*{Power spectral density}
In a practical implementation, the output signal of the homodyne interferometer $I_H(t)$ is handled by an electronic spectrum analyzer which measures the spectrum of fluctuations $\delta I_H(t)$ and quantifies these in terms of a (symmetrized) power spectral density. 

We note that if two variables are linearly (and instantaneously) related via $x=ay$, their spectral densities are related through $S_{xx}=a^2S_{yy}$. If we assume that the amplitude of the modulation $\delta\omega_c(t)$ is much smaller than the cavity linewidth $\kappa$, the variation of $I_H$ can be approximated by the Taylor expansion truncated to first order: $\delta I_H(t)=\frac{\partial\bar{I}_H}{\partial \bar{\omega_c}}\delta\omega_c(t)$, as we did in equation~(\ref{eq:ph_fluc}).
In this case it is possible to write an analytical expression for the transduction from a given fluctuation spectrum of $\omega_c$ to $I_H$: the spectral density of the frequency fluctuations $S_\mathrm{\omega\omega}$ is related to the measured signal from the spectrum analyzer $S_ {I_HI_H}$, by means of the relationship \cite{Gorodetsky2010}
\begin{equation}\label{eq:betadef}
    S_ {I_HI_H} (\Omega) = \frac{\beta^2}{\kappa^2} \frac{P_\mathrm{in}P_\mathrm{lo}}{(\hbar\omega_l)^2} S_\mathrm{\omega\omega} (\Omega),
\end{equation}
where the \textit{transduction function} $\beta$ was introduced in equation~(\ref{eq:ph_fluc})
\begin{eqnarray} 
    \beta^2(\bar{\Delta}_n,\eta,\theta) &=& \frac{64 \eta^2}{(1+\bar{\Delta}_n^2)^4} \nonumber \\ 
    && \times \left[(1-\bar{\Delta}_n^2)\cos(\theta)-2\bar{\Delta}_n\sin(\theta) \right]^2
\label{eq:beta2}
\end{eqnarray}
and we assume $\theta = \theta(\bar{\Delta}_n,c_n,\varphi)$ due to the way the homodyne is locked.

\begin{figure}
    \centering
    \includegraphics[width=0.48\textwidth]{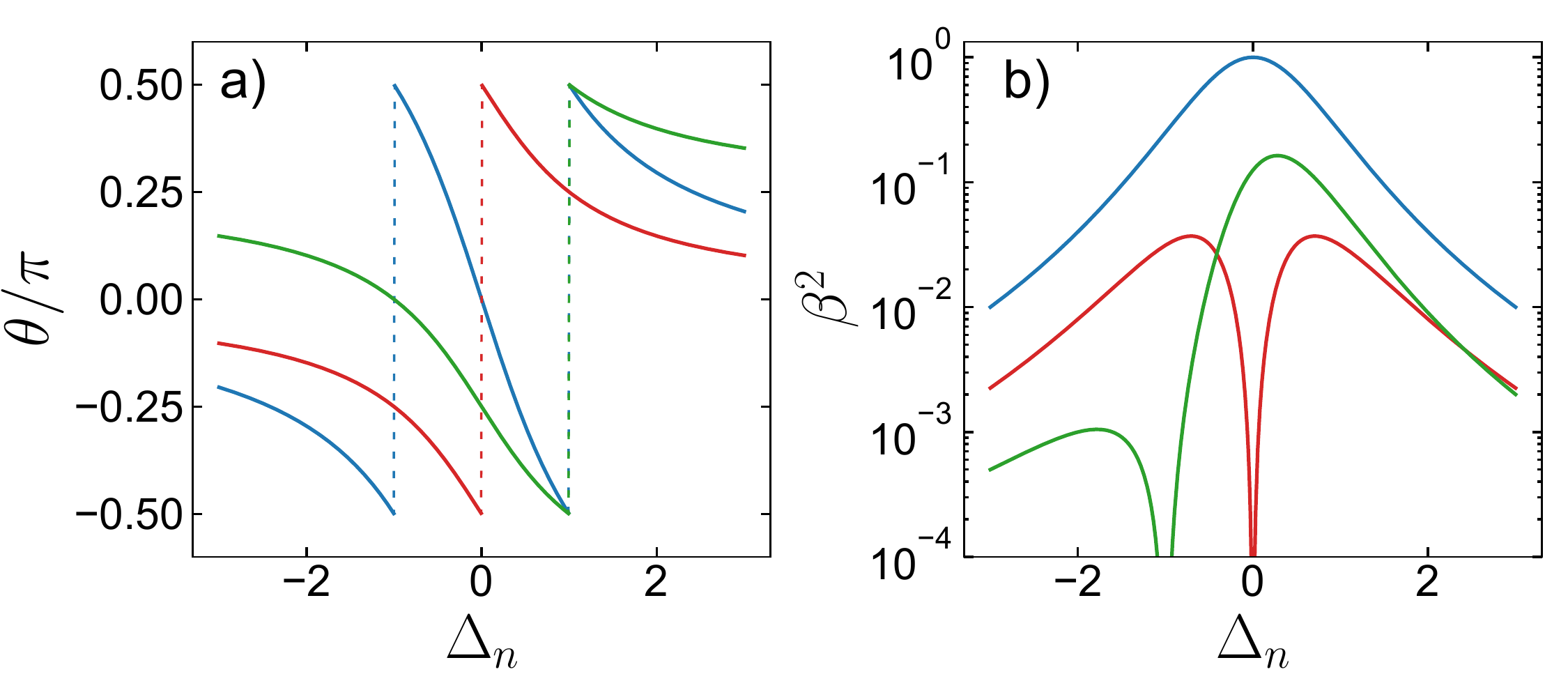}
    \caption{Transduction function versus laser detuning when $|\bar{I}_H|$ is minimized. Panel a) shows the effective homodyne angle and panel b) the $\beta^2$ parameter. The different colors correspond to the three cavities shown in figure \ref{fig:fpcavity}: single sided cavity (blue), double sided cavity (red), and double sided cavity with a non-resonant channel (green).}
    \label{fig:3}
\end{figure}

\begin{figure}
    \centering
    \includegraphics[width=0.48\textwidth]{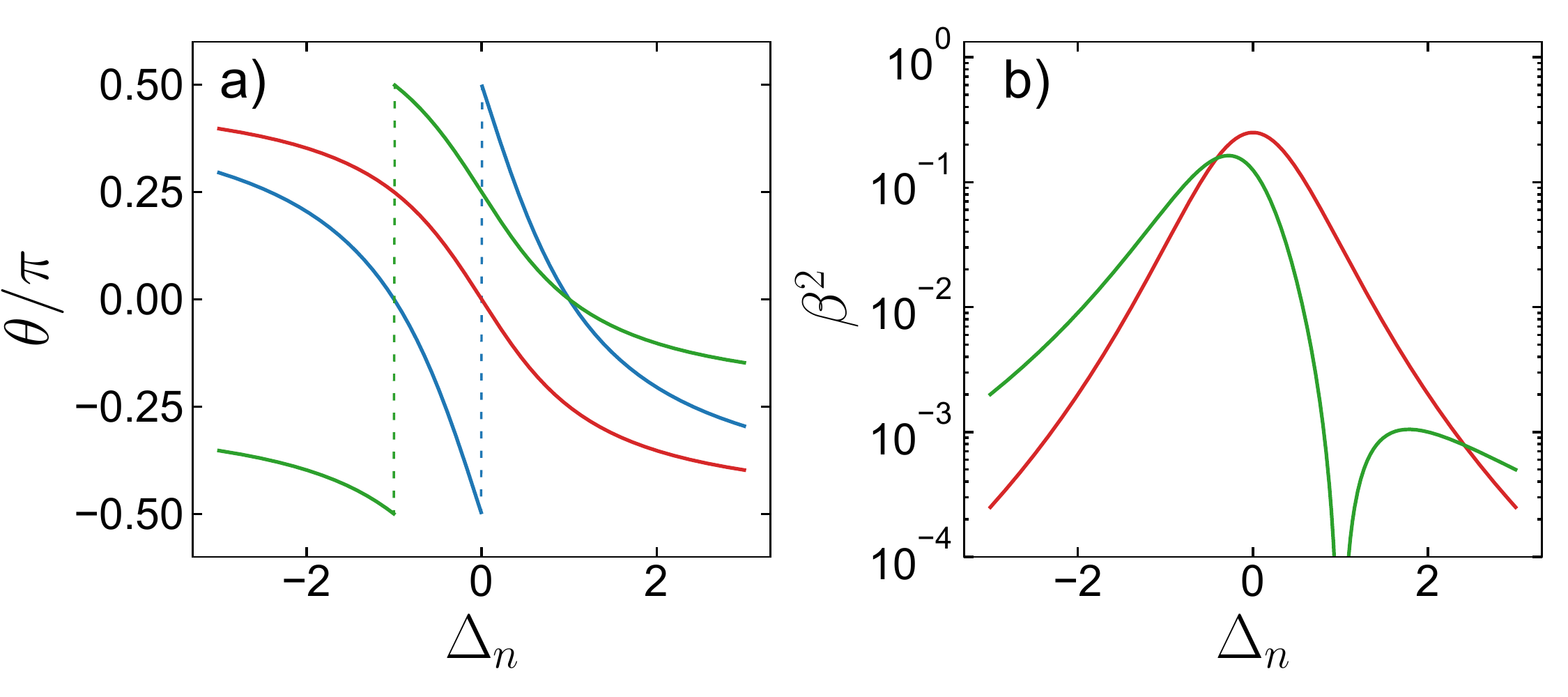}
    \caption{Transduction function versus laser detuning when $|\bar{I}_H|$ is maximized. Panel a) shows the effective homodyne angle and panel b) the $\beta^2$ parameter. The different colors correspond to the three cavities shown in figure \ref{fig:fpcavity}: single sided cavity (blue), double sided cavity (red), and double sided cavity with a non-resonant channel (green). Note that the single sided cavity $\beta$ is now identically zero and hence not shown.}
    \label{fig:4}
\end{figure}

Figures \ref{fig:3} and \ref{fig:4} show the transduction function $\beta^2$ for the three types of cavities depicted in figure~\ref{fig:fpcavity}, with the assumption that the homodyne is locked by either minimizing (figure~\ref{fig:3}) or maximizing (figure~\ref{fig:4}) $|I_H|$. All figures have been normalized by dividing equation~(\ref{eq:beta2}) by 64 (so that the maximum with $\eta=1$ is one). From these figures, we see that the shape of the transduction function can vary considerably depending on the non-resonant parameters, making the experimental results hard to interpret, especially if $c$ and $\varphi$ are not well controlled. 

\section{Quadrature-averaged homodyne detection}
We now show that introducing a periodic modulation on the reference beam phase leads to a result that is closer to the ideal homodyne detection, from the point of view that the dependence of $\theta$ on the non-resonant scattering parameters is erased. This in turn implies that the measurement probes only the resonant channel of the optical system, described by the second term of equation~(\ref{eq:tf}). %The \textit{quadrature averaged} homodyne detection sees all the considered optical system as characterized by a Lorentzian resonance of the optical mode. 

We assume that the phase difference between the signal and local oscillator is now harmonically modulated in time $\theta=2\pi \nu t$ and that the time of acquisition is sufficiently large (meaning that the signal is collected over multiple periods of modulation), so that we can average $\left<\delta I_H^2\right>$ over time. Using equation~(\ref{eq:beta2}) and 
\begin{align}
    &\int_\tau dt \cos^2(2\pi\nu t)=\frac{1}{2\pi}{\int_0}^{2\pi}  d\theta \cos^2\theta= \frac{1}{2} \nonumber \\
    &\int_\tau dt \sin^2(2\pi\nu t)=\frac{1}{2\pi}{\int_0}^{2\pi}  d\theta \sin^2\theta= \frac{1}{2} \nonumber \\
    &\int_\tau dt \sin(2\pi\nu t)\cos(2\pi\nu t)=\frac{1}{2\pi}{\int_0}^{2\pi}  d\theta\sin{\theta}\cos\theta=0, \nonumber
\end{align}
we arrive at an expression for the transduction of the form
\begin{equation}
    \boxed{
    \beta_\mathrm{QA}^2 (\bar{\Delta}, \eta)= \frac{32 \eta^2}{\left(1+\bar{\Delta}_n^2\right)^2}
    }.
\label{eq:betaQA}
\end{equation}

\begin{figure}
    \centering
    \includegraphics[width=0.48\textwidth]{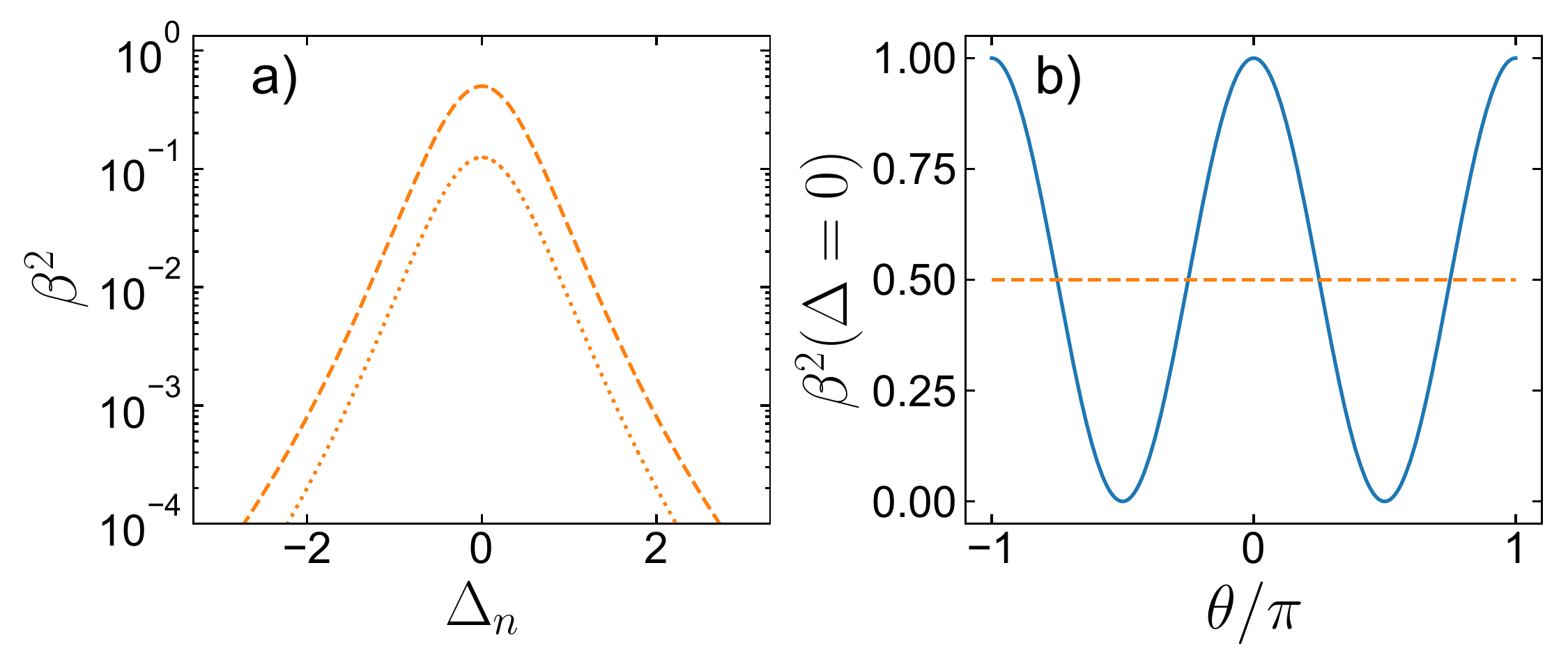}
    \caption{a) Transduction function for quadrature averaged homodyne detection. The dashed line assuming $\eta=0.5$ and the dotted line with $\eta = 1$. b) Comparison between transduction function for single quadrature and quadrature averaged measurements versus angle $\theta$ for $\bar{\Delta}=0$, when $\eta=1,c=1,\varphi=0$. At every angle the quadrature averaged measurement is at most factor of 2 less sensitive.}
    \label{fig:QA}
\end{figure}

The transduction now has a simple dependence on the cavity resonance parameters and by fitting the experimental data across all laser frequencies with two parameters it is possible to characterize the cavity resonance. In figure~\ref{fig:QA} we show the $\beta_\mathrm{QA}$ that now assumes the same detuning dependence for all the cases depicted in figure~\ref{fig:fpcavity}. This has to be compared with the result of figures \ref{fig:3} and \ref{fig:4} already discussed.

The drawback of this method is a decreased sensitivity but that amounts to maximum a factor of two with respect to the ideal case.
In figure~\ref{fig:QA} we show that the sensitivity of the measurement in the quadrature averaged case is indeed decreased by at most a factor of 3 dB. In order to do so we compare the case of a single sided Fabry-Pérot cavity for which the sensitivity is maximum, at $\Delta=0$. The maximum sensitivity that follows from equation~(\ref{eq:beta2}) is
\begin{equation}
    \beta(\bar{\Delta}=0)= 64 \eta^2\cos^2\theta
    =2 \cos^2\theta\beta_\mathrm{QA}(\Delta=0),
\end{equation}
from which it follows that the decrease in the sensitivity happens for $\theta=n\pi$ (with $n$ integer), that corresponds to phase quadrature.

One could notice that the way we operate the homodyne interferometer here would correspond to performing an heterodyne detection with a negligible frequency difference between the two beams of the interferometer, hence we could call this technique ``DC-heterodyne detection''. Hence, the decreased sensitivity compared to a ``pure'' homodyne configuration can be explained as a fundamental limitation deriving from signal to noise limitation of a heterodyne detector.

\section{Characterization of absolute magnitude of modulation}
In the previous section we showed that, by averaging the output signals over the different field quadratures, we obtain a simple expression for the transduction $\beta$, that describes how the fluctuations of a cavity frequency transform the output signal of our homodyne interferometer. Importantly, we made the approximation that the amplitude of the cavity frequency modulation $\delta\omega$ is small compared to the cavity linewidth $\kappa$ and we considered only the first term in the Taylor expansion of the output signal $I_H$. Our function $\beta_\mathrm{QA}$ is relative to the \textit{linear transduction} of the modulation $\delta\omega$ since it transforms a modulation signal on the cavity frequency in the homodyne output, in a linear fashion. By retaining only the linear term, we discarded the nonlinear part of $I_H$, losing precious information about our system. Now, while considering the case of harmonic modulation at a frequency $\Omega$: $\delta\omega=A\cos(\Omega t)$, we show how one can use the information contained in nonlinear part of $I_H$ in order to make predictions on the modulation amplitude $A$ \cite{leijssen_nonlinear_2017}.
    
We can write the full Taylor expansion of $I_H(t)$ as
\begin{eqnarray}
    I_H(t) &=& \sum_{k=0}^\infty \frac{1}{k!} \frac{\partial^k I_H(\bar{\Delta})}{\partial\Delta^k} [\delta\omega(t)]^k \nonumber \\
    &=& \sum_k \frac{1}{k!} \frac{\partial^k I_H}{\partial \Delta^k} \left[A\cos(\Omega t)\right]^k.
\end{eqnarray}
Since we still want to consider the case of small modulation amplitudes: $A\ll\kappa$, it is possible to make the following ``order by order'' approximation \cite{Hauer2015,brawley_nonlinear_2016,leijssen_nonlinear_2017}
\begin{equation}
    \cos^k(\Omega t) \simeq \frac{1}{2^{k-1}}\cos(k\Omega t),
\label{eq:orderapprox}
\end{equation}
since, when $A$ is small, the convergence of the sum is ensured by cutting the sum index to a finite value. The relationship in equation~(\ref{eq:orderapprox}) makes it possible to immediately write $I(t)$ as a Fourier series
\begin{equation}
    I_H(t)=
    \sum_k \frac{1}{k!} \frac{\partial^k I_H}{\partial \Delta^k} \frac{A^k}{2^{k-1}}\cos(k\Omega t) = \sum_k I_k(t).
\label{eq:fourier}
\end{equation}
At this point one can exploit the same method already used for the derivation of $\beta_\mathrm{QA}$, in order to define an expression of the transduction correspondent to each $k$-th order in $\Omega$, $\beta_\mathrm{QA,k}$. In particular, the operator  $\beta_\mathrm{QA,k}$ is defined by an analogous relationship as in (\ref{eq:betadef}), by noticing that every term $I_k(t)$ is linear in $\cos(k\Omega t)$. 
   
In order to derive an expression for the transduction, we need to evaluate $\beta^2_\mathrm{QA,k} = \left<\left[\frac{\partial^{k} I_H}{\partial \Delta^{k}}\right]^2\right>_\theta$
\small
\begin{eqnarray}
&&\left<\left[\frac{\partial^{k} I_H}{\partial \Delta^{k}}\right]^2\right>_\theta = (-4 \eta \left|s_\mathrm{in} \right|  \left|s_\mathrm{lo}\right|)^2 \nonumber \\
&\times& \left<\left[\frac{\partial^{k}}{\partial \Delta^{k}} \left(\frac{\Delta_n}{1+\Delta_n^2}\cos\theta - \frac{1}{1+\Delta_n^2} \sin\theta\right) \right]^2\right>_\theta \\
&=& 8 (\eta \left|s_\mathrm{in} \right|  \left|s_\mathrm{lo}\right|)^2 \left[\left(\frac{\partial^{k}}{\partial \Delta^{k}}  \frac{\Delta_n}{1+\Delta_n^2}\right)^2 + \left(\frac{\partial^{k}}{\partial \Delta^{k}}  \frac{1}{1+\Delta_n^2}\right)^2 \right] \nonumber
\end{eqnarray}
\normalsize
where in the last line we have done the averaging over all angles $\theta$ as previously.

For simplifying notation, we mark $f = \Delta_n/(1+\Delta_n^2)$, $g = 1/(1+\Delta_n^2)$ and $\partial^k = \partial^k/\partial \Delta^{k}$. We then have to solve for $(\partial^k f)^2$ and $(\partial^k g)^2$.
Since they are both real valued we can write
\begin{align}
    (\partial^k f)^2+(\partial^k g)^2=(\partial^kf+i\partial^kg)(\partial^kf-i\partial^kg).
\label{eq:square}
\end{align}
At this point it is possible to write an expression for the $k$-th order derivative of both $f$ and $g$. Indeed, in view of linearity of the derivative we can write
\begin{eqnarray}
    \frac{\partial^k}{\partial \Delta^k}(f+ig) &=& \frac{\partial^k}{\partial \Delta^k}\left(\frac{\Delta_n+i}{1+\Delta_n^2}\right) \nonumber \\
    &=& \frac{\partial^k}{\partial \Delta^k}\frac{1}{\Delta_n-i} \nonumber \\
    &=&\left(\frac{-2}{\kappa}\right)^k \frac{k!}{(\Delta_n-i)^{k+1}}   \nonumber
\end{eqnarray}
and in the same way
\begin{equation*}
    \frac{\partial^k}{\partial \Delta^k}(f-ig)=\left(\frac{-2}{\kappa}\right)^k \frac{k!}{(\Delta_n+i)^{k+1}}. 
\end{equation*}

Equation~(\ref{eq:square}) then becomes
\begin{equation*}
    (\partial^k f)^2+(\partial^k g)^2=\left(\frac{-2}{\kappa}\right)^{2k} \frac{(k!)^2}{(\Delta_n^2+1)^{k+1}},
\end{equation*}
and this ultimately implies that
\begin{equation}
    \left<\left[\frac{\partial^{k} I_H}{\partial \Delta^{k}}\right]^2\right>_\theta =8\eta^2 \frac{P_\mathrm{in}P_\mathrm{lo}}{(\hbar\omega_l)^2} \left(\frac{2}{\kappa}\right)^{2k}  \frac{(k!)^2}{(\Delta_n^2+1)^{k+1}}.
\end{equation}
    
We can call $S_{I_kI_k}$ the PSD of the $k$-th term in equation~(\ref{eq:fourier}) and $S_\mathrm{\Omega_k\Omega_k}$ the delta function correspondent to the $\cos(k\Omega t)$, so that
\begin{eqnarray}
    S_{I_kI_k} &=& \beta^2_{\mathrm{QA,}k} \frac{P_\mathrm{in}P_\mathrm{lo}}{(\hbar \omega_l)^2} S_\mathrm{\Omega_k\Omega_k} \nonumber \\
    &=& \beta^2_{\mathrm{QA,}k} \frac{P_\mathrm{in}P_\mathrm{lo}}{(\hbar \omega_l)^2} \delta(\Omega-\Omega_k).
    \label{eq:defbetak}
\end{eqnarray}
Note that we have now included the $\kappa$ term inside $\beta$ unlike in equation~(\ref{eq:beta2}). From equations (\ref{eq:fourier}) and (\ref{eq:defbetak}) it then follows that
\begin{equation}
    \boxed{
    \beta^2_{\mathrm{QA,}k} =\left(\frac{A}{\kappa}\right)^{2k}\frac{32\eta^2}{(1+\Delta_n^2)^{k+1}}}.
    \label{eq:beta_obo}
\end{equation}

From equation~(\ref{eq:beta_obo}) one can see that by comparing the relative weights of the different harmonic components we can arrive at an estimate of $A$ that is independent of constant terms such as the powers $P_\mathrm{in},P_\mathrm{lo}$ or the coupling factor $\eta$. For example if we look at the ratio of the higher order modulation with respect to the lower order at zero detuning
\begin{equation}
    \frac{S_{I_{k+1}I_{k+1}}}{S_{I_kI_k}}\bigg|_{\bar{\Delta}=0}=\frac{\beta_{\mathrm{QA,}k+1}}{\beta_{\mathrm{QA,}k}}\bigg|_{\bar{\Delta}=0}=\left(\frac{A}{\kappa}\right)^2,
\label{eq:ratiopeaks}
\end{equation}
we see that there is a straightforward connection to the \textit{modulation strength} defined as the ratio between the modulation amplitude and the cavity linewidth $A/\kappa$. This is enforced by the fact that contrarily to the standard case, in the quadrature-averaged case, the transduction peaks at $\Delta=0$ at all the orders.

For large modulations $A\simeq\kappa$, the approximation in (\ref{eq:orderapprox}) is no longer valid, and one cannot exploit this method.  In this case it is only possible to give a numerical estimate of the shape of the transduction operator for the higher orders since we cannot anymore exploit the convergence of the truncated Taylor expansion valid for small perturbations. 

\section{Experimental demonstration}
\begin{figure}
    \centering
    \includegraphics[width=0.48\textwidth]{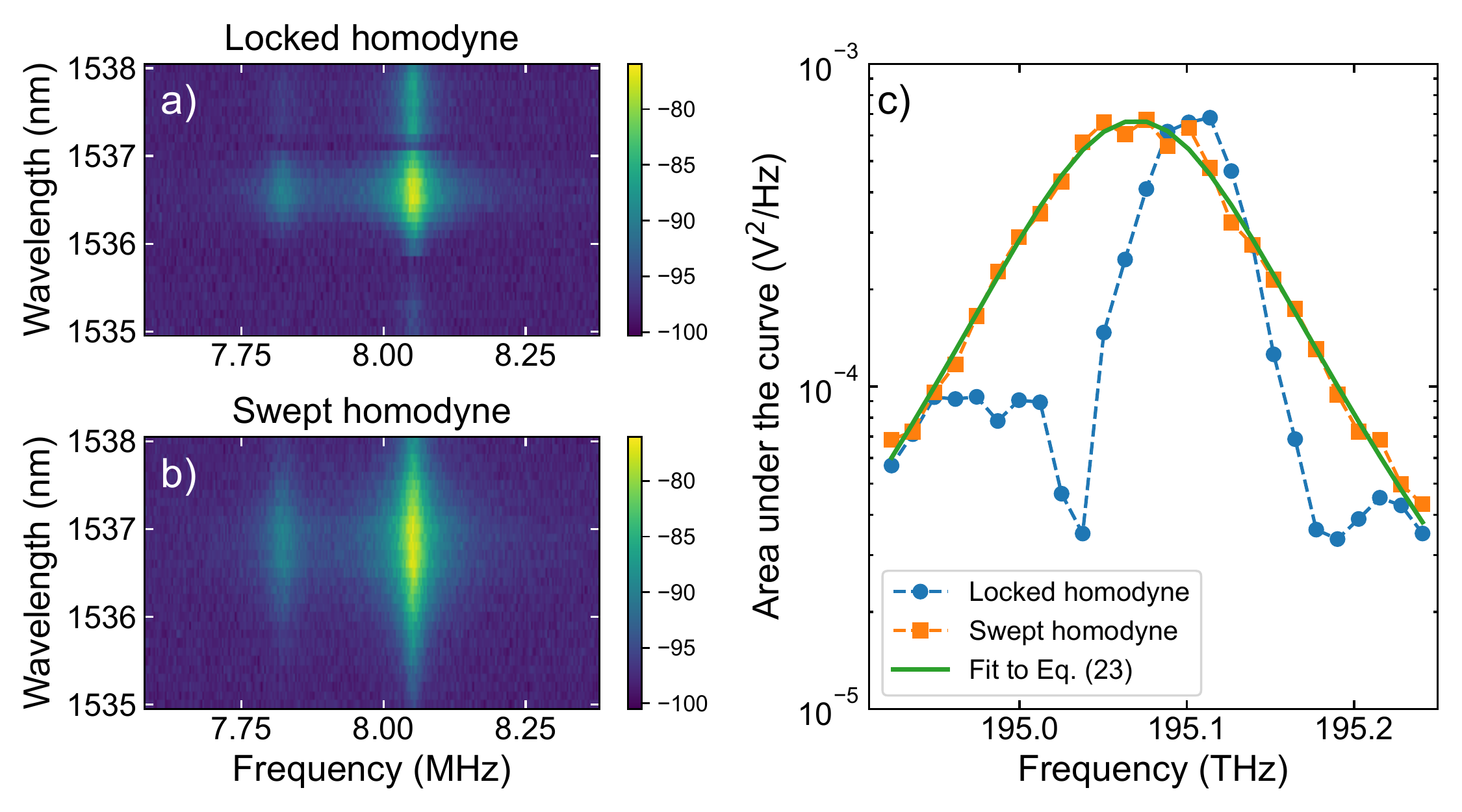}
    \caption{Spectrograms for the locked homodyne (a) and swept homodyne (b) measurement for the first order mechanical signal. The colorscale shows the measured voltage signal ($I_H$ after a transimpedance amplifier) in dB scale. Panel c) shows the integrated area under the mechanical signal peak as a function of laser detuning, together with a fit to the equation~(\ref{eq:betaQA}) for the swept homodyne case. The lower frequency signal is another mechanical mode that we do not consider here.}
    \label{fig:experiment1}
\end{figure}

To compare the above formulation to experiments, we have performed measurements on a sliced photonic crystal nanobeam resonator \cite{leijssen_strong_2015,leijssen_nonlinear_2017} using both the ``locked homodyne'' method (with absolute value of the DC photocurrent minimized)  and the ``swept homodyne'' method presented above. The results are analyzed in figure~ \ref{fig:experiment1} where the area under the mechanical resonance peak is plotted as a function of the laser wavelength, together with the bare spectrograms. The experiments are performed in a room temperature vacuum chamber (where the temperature and vacuum pressure are constant) with a focused laser beam incident from free space at normal incidence to the sample and measured in reflection, and with a constant laser power so that the variation in the measured signal as a function of the wavelength is solely due to variation of the transduction parameter $\beta$ in equation~(\ref{eq:betaQA}). As expected, our data shows that whereas the swept homodyne case produces a well defined single resonance from which the optical cavity properties can be extracted ($\kappa/2\pi \approx 192$ GHz), the locked homodyne shows a distinct Fano shape, that is significantly harder to interpret and can even lead to a significant underestimation for the cavity linewidth if not properly analyzed.

\begin{figure}
    \centering
    \includegraphics[width=0.48\textwidth]{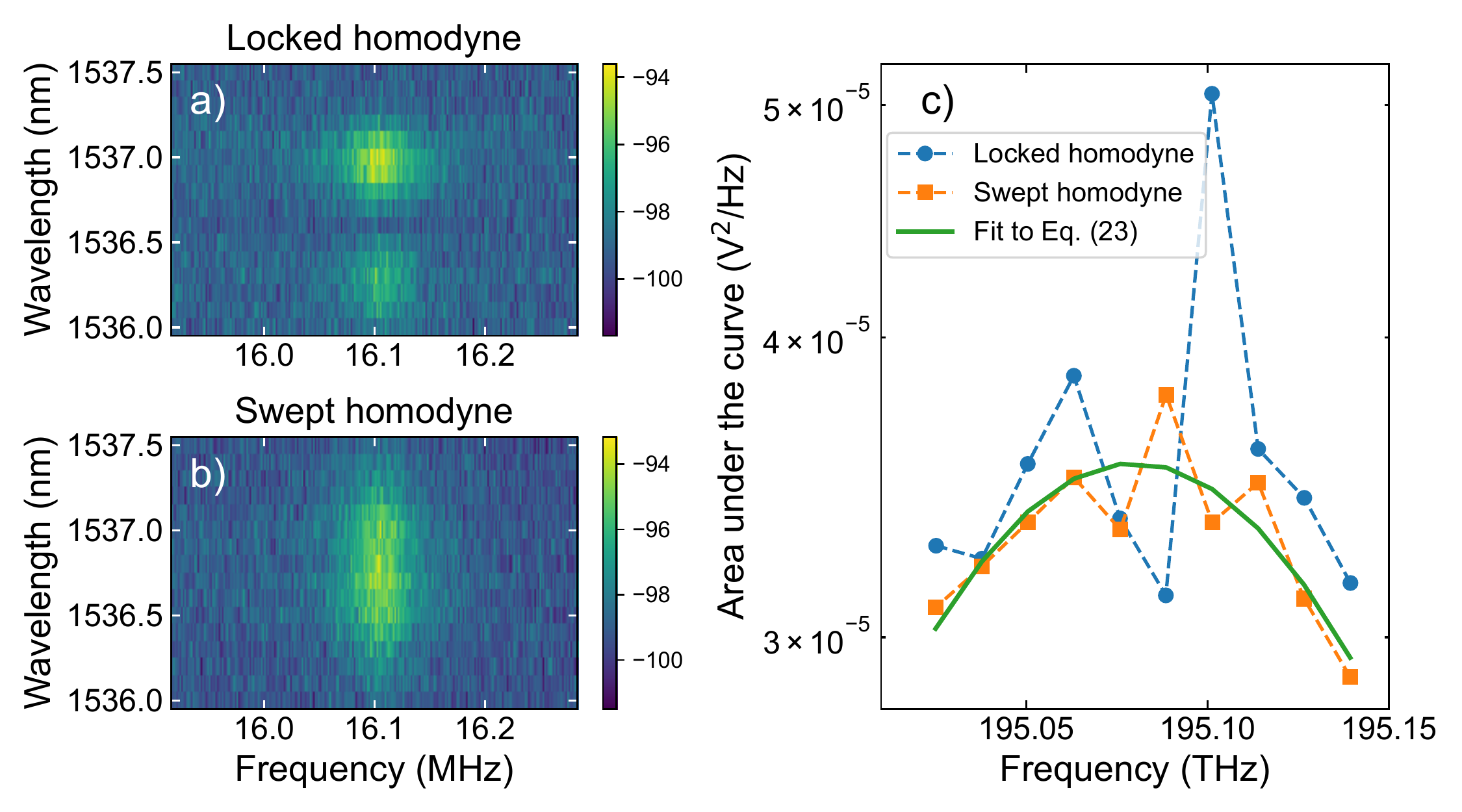}
    \caption{Spectrograms for the locked homodyne (a) and swept homodyne (b) measurement for the second order mechanical signal. The colorscale shows the measured voltage signal ($I_H$ after a transimpedance amplifier) in dB scale. Panel c) shows the integrated area under the mechanical signal peak as a function of laser detuning, together with a fit to the equation~(\ref{eq:betaQA}) for the swept homodyne case.}
    \label{fig:experiment2}
\end{figure}

In figure~\ref{fig:experiment2} similar data is presented for the second order peak. Similarly as in the first order case, sweeping the homodyne phase turns the Fano shape into a simple peak. More interestingly, from this data we can extract the ratio of the signal from second order peak to the first order, which turns out to be $\sim$0.053. This corresponds to ratio $A/\kappa$ of 0.23 according to equation~(\ref{eq:beta_obo}). Assuming that the amplitude of the frequency fluctuations follows $A=2\delta \omega_\mathrm{RMS} = 2\sqrt{2n_{th}}g_0$ (where $n_{th} = k_BT/(\hbar \omega_m))$ and using $\kappa/2\pi = 192$ GHz extracted above, we get that $g_0/2\pi \approx 18$ MHz, which is is well in line with the parameters we have extracted before for similar resonators using independent methods.

\section{Conclusions and discussion}
In conclusion, we have shown both theoretically and numerically that averaging a homodyne interferometer over all possible measurement angles can have advantages in specifically estimating the parameters of resonant cavities. We think the fact that the homodyne interferometer angle can depend very sensitively on the detuning and non-resonant parameters is an underappreciated feature that should be considered carefully in experiments. Fortunately, the averaging presented here can remedy this problem.

\begin{acknowledgements}
We thank Amy Navarathna for assistance in sample fabrication. This project has received funding from the European Research Council (ERC) under the European Union’s Horizon 2020 research and innovation programme (grant agreement No 852428) and from Academy of Finland Grant No 321416. This work is part of the research programme of the Netherlands Organisation for Scientific Research (NWO), and supported by an NWO Vidi grant.
\end{acknowledgements}

\bibliography{swhd_refs}

\end{document}